\PassOptionsToPackage{twoside=false}{geometry}
\documentclass[sigconf, nonacm]{acmart}

\usepackage{graphicx}
\usepackage{booktabs}
\usepackage{wrapfig}
\usepackage{subcaption}
\usepackage{makecell}
\usepackage{listings}
\usepackage{xspace}

\AtBeginDocument{%
  }

\newcommand{\tallm}{\textsc{TaLLM}\xspace}
\newcommand{\tallms}{\textsc{TaLLM}s\xspace}
\def\taubench{$\tau$-bench }

\begin{document}

\title{Solver-Aided Verification of Policy Compliance in Tool-Augmented LLM Agents}

\author{Cailin Winston}
\affiliation{%
  \institution{University of Washington}
  \city{Seattle}
  \country{USA}
}

\author{Claris Winston}
\affiliation{%
  \institution{University of Washington}
  \city{Seattle}
  \country{USA}
}
\author{Ren\'e Just}
\affiliation{%
  \institution{University of Washington}
  \city{Seattle}
  \country{USA}
}

\begin{abstract}
Tool-augmented Large Language Models (\tallms) extend LLMs with the ability to invoke external tools, enabling them to interact with real-world environments. However, a major limitation in deploying \tallms in sensitive applications such as customer service and business process automation is a lack of reliable compliance with domain-specific operational policies regarding tool-use and agent behavior. Current approaches merely steer LLMs to adhere to policies by including policy descriptions in the LLM context, but these provide no guarantees that policy violations will be prevented. In this paper, we introduce an SMT solver-aided framework to enforce tool-use policy compliance in \tallm agents. Specifically, we use a LLM-assisted, human-guided approach to translate natural-language-specified tool-use policies into formal logic (SMT-LIB-2.0) constraints over agent-observable state and tool arguments. At runtime, planned tool calls are intercepted and checked against the constraints using the Z3 solver as a pre-condition to the tool call. Tool invocations that violate the policy are blocked. We evaluated on the \taubench benchmark and demonstrate that solver-aided policy checking reduces policy violations while maintaining overall task accuracy. These results suggest that integrating formal reasoning into \tallm execution can improve tool-call policy compliance and overall reliability.

\end{abstract}

\maketitle

\begin{figure}[]
\begin{center}
\includegraphics[width=\linewidth]{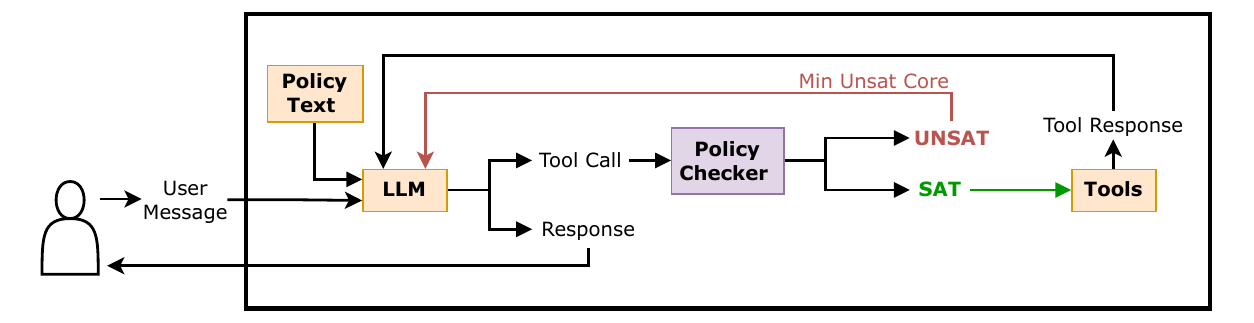}
\end{center}
\caption{Tool-augmented LLM with tool-call policy checker. The framework integrates an SMT-based policy checker into the tool-call execution loop, blocking tool invocations that violate specified tool-use policies.}
\label{fig:architecture}
\end{figure}

\section{Introduction}

Tool-augmented Large Language Models (\tallms) enable LLMs to invoke external tools that allow them to interact with and act in real-world environments. Such advancements have accelerated the deployment of LLMs in autonomous agents. Tool planning and tool use capabilities allow LLM agents to retrieve up-to-date information and autonomously take actions beyond text responses. For example, a customer service agent for an airline company may be backed by an LLM enabled to invoke tools to query a user database and to book or cancel flight reservations. While this capability increases autonomy, it also introduces risks around the correctness and safety of actions taken, especially in domains where strict adherence to a set of operational policies is critical. Incorrect or unauthorized tool calls may lead to policy violations or irreversible consequences. Hence, ensuring that tool calls invoked by the LLM agent comply with domain-specific tool-use policies prior to execution is critical for safe and reliable deployment.

Although much prior work has focused on improving the tool planning and usage capabilities of TaLLMs as well as the overall end-to-end task accuracy, safeguarding against policy violations remains an open question. Existing policy enforcement mechanisms for TaLLM agents typically involve embedding long policy documents in the system prompt or applying output guardrails based on heuristic or pattern-matching rules. These approaches lack formal semantics and do not provide reliable guarantees that policy violations will be prevented. One could also integrate a human trained with domain expertise in the loop to authorize each tool call before execution. However, this limits agent autonomy and processing speed as it requires the human to be familiar with the context of each request. Furthermore, all these approaches are not scalable to larger and more complex policies.

Focusing on the set of policies that directly govern and constrain tool use, we argue that tool-call policy compliance can be modeled as a constraint satisfaction problem over agent-observable state and the tool set and can be implemented as a runtime check before tool execution. With this framework, determining whether a tool call violates a policy reduces to checking the satisfiability of a set of logical constraints derived from the policy, the agent's current state, and the planned tool call.

We propose a formal-methods-based framework that encodes tool-use policies as logical constraints and checks tool calls for compatibility prior to tool execution using an SMT solver. By verifying planned actions before they occur, our method aims to reduce the proportion of policy-violating tool calls while maintaining task success rates. Figure~\ref{fig:architecture} illustrates such a TaLLM agent integrated with a SMT-solver-based policy checker.

Tool-use policies are often defined in long natural-language policy documents that contain both tool-use policies and other behavioral or operational policies that are not tied to specific tool-use. While translating such documents into formal logic constraints is an important challenge, our primary focus in this work is evaluating the effectiveness of using formal logic if such a translation exists. Hence, we extract tool-use policies from policy documents and encode them into formal logic statements using an LLM-assisted, human-guided, and human-reviewed approach. This allows us to isolate evaluating the effectiveness of constraint-based runtime enforcement of tool-use policies.

Specifically, we make the following contributions:
\begin{itemize}
    \item A formalization of TaLLM agent tool-use policies as logical constraints over agent-observable state and tool arguments.
    \item SMT solver-based policy compatibility checking framework for TaLLMs.
    \item An empirical evaluation on the $\tau^2$-bench, demonstrating the effectiveness and feasibility of solver-aided policy checking for TaLLMs to reduce policy-violating tool calls.
\end{itemize}

\section{Logical Encoding of Natural Language Policies}

Operational policies for TaLLM agents are typically specified in natural language documents that may be derived from real-world policy documents for the domain in which the agent is deployed. For example, a policy for an airline customer service agent might contain response guidelines and rules describing when operations such as cancellations and reservation changes are permitted. Since LLMs operate on natural language, these policy documents can be provided as text inputs to the LLM to attempt to steer the LLM to adhere to them.

However, prompt-based guidance does not provide reliable policy enforcement. As policies grow longer and more complex, purely prompt-based enforcement becomes increasingly brittle and harder to validate. LLMs may struggle with remember complex policies and their interactions across multi-turn conversations and across various tool invocations. Furthermore, there is no way to validate that a proposed tool call satisfies all policy requirements before execution.

We observe that tool-use policies are fundamentally constraint systems: they specify conditions over an agent's observable state and tool arguments that determine whether a tool call is permitted. This observation enables modeling policy compliance as a satisfiability problem. We first model a policy as a set of unconstrained variables representing all relevant state and possible tool calls along with predicates capturing constraint-based relationships between the variables. Then, given a proposed tool call and the agent’s observable state that constrain a subset of all policy variables, do there exist assignments to the remaining unconstrained variables that satisfy all policy constraints if the tool were called? If not, the tool call violates the policies. Once encoded as SMT formulas, a solver such as Z3 can automatically determine satisfiability.

\subsection{Policy Encoding}

Given a set of policies in natural language, a domain expert converts them to a logical encoding that contains:

\begin{itemize}
    \item \textbf{Uninterpreted constants} representing observable state and tool arguments (e.g., user identifiers, timestamps, flags indicating whether required information has been obtained).
    \item \textbf{Sorts} capturing enum-like categories (e.g., membership tiers, reservation states).
    \item \textbf{Predicates and functions} implemented as uninterpreted constants but semantically encoding policy conditions (e.g., whether a cancellation is allowed).
    \item \textbf{Assertions} that constrain relationships among these constants and predicates.
\end{itemize}

\subsection{Tool Validation Schema}

In order to enable reliable policy checking at runtime, a domain expert defines a validation schema for each tool available to the LLM agent that is based on the policy encoding. A schema specifies (1) the set of variables from the policy encoding that must be instantiated to assess compatibility for that tool, including their types and any domain-specific enumerations, and (2) a designated policy predicate whose satisfiability determines whether the tool invocation is permitted. Tools that are unconditionally allowed are associated with empty schemas and bypass compatibility checking. An example of a policy encoding and its corresponding tool schema for a cancellation tool is showed in Figure~\ref{fig:checker_flow}.

\section{Solver-Aided Policy Compatibility Checking}

\begin{figure*}[t]
\begin{center}
\includegraphics[width=\linewidth]{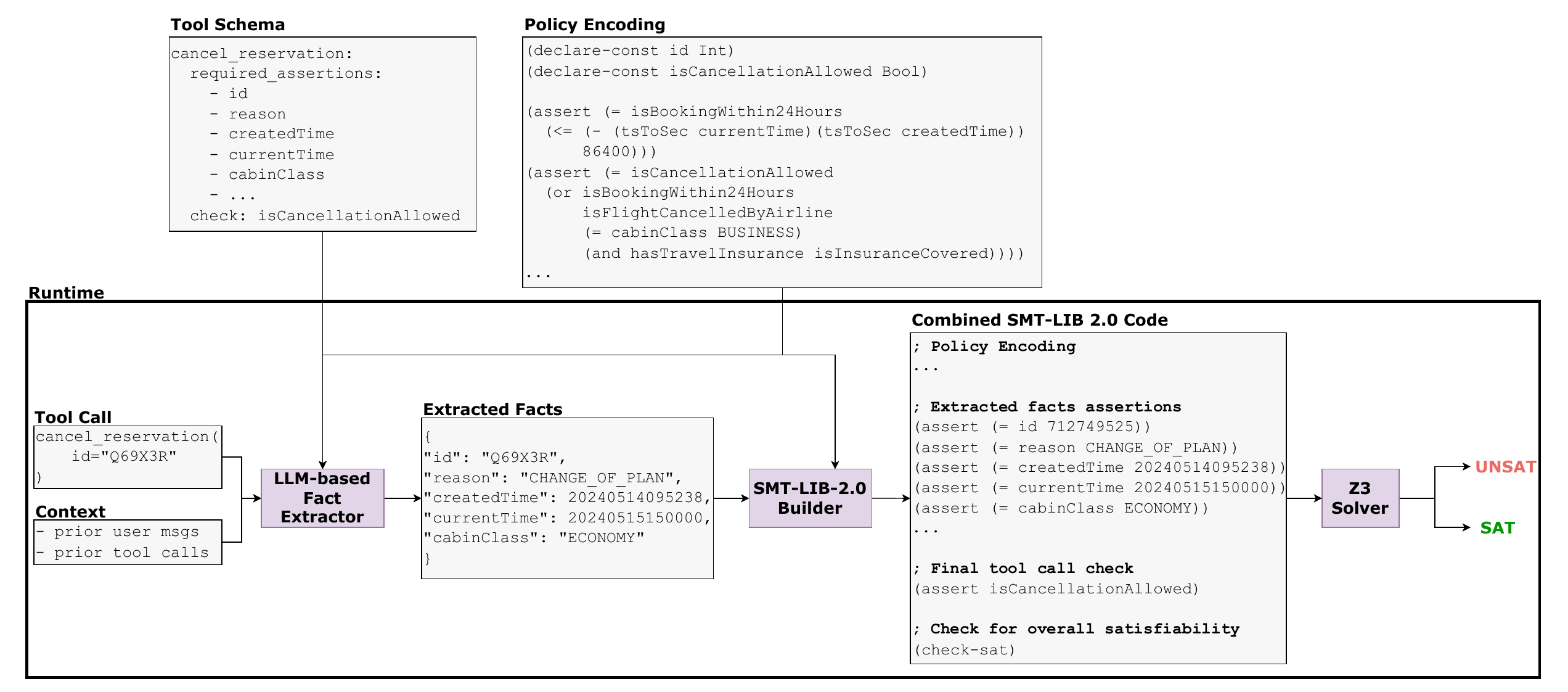}
\end{center}
\caption{Flow diagram of the policy compatibility checker in order to validate whether a tool call for cancellation can be called.}
\label{fig:checker_flow}
\end{figure*}

Once a logical encoding of tool-use policies is constructed, these constraints can be applied at runtime to verify tool-call compatibility. We introduce a solver-aided policy compatibility checker that is integrated into the tool planning loop of a TaLLM. At runtime, the checker intercepts each planned tool call, constructs a concrete set of logical assertions using the agent’s observable state and planned tool call, and checks for satisfiability against the policy encoding using a SMT solver. The solver’s result is then used to allow or block the tool invocation prior to execution. Figure~\ref{fig:checker_flow} depicts an example of policy checking works for a single tool call.

\subsection{Runtime Tool-Call Interception}

At runtime, the policy compatibility checker is invoked whenever the agent produces a candidate tool call. The checker observes the tool name, its arguments, and the preceding conversation history, but does not otherwise alter the agent’s internal reasoning process. Instead, it acts as an external gate that validates whether the proposed tool invocation is compatible with the policy encoding under the current agent-observable state.

\textbf{SMT-LIB Constraint Construction} Given a planned tool call, the checker retrieves the corresponding tool validation schema and instantiates the required policy variables. A language model is used to extract concrete values for these variables from the conversation history and the tool arguments.

The extracted values are translated into logical assertions and appended to the static SMT-LIB policy encoding. The designated policy predicate according to the tool schema is then asserted as the compatibility condition to be checked for the current tool invocation.

\textbf{Solver-Based Policy Checking} The constructed SMT-LIB code is submitted to an SMT solver, which determines whether it is satisfiable (\texttt{SAT}) or unsatisfiable (\texttt{UNSAT}). A \texttt{SAT} result indicates that there exists an assignment consistent with both the policy constraints, the current state, and the proposed tool call, and the invocation is permitted. A \texttt{UNSAT} result indicates a policy violation, and the tool call is blocked.

\subsection{Agent Planning Control}

If the solver returns \texttt{SAT}, the tool call is executed. If it returns \texttt{UNSAT}, the tool call is blocked and the agent is prompted to revise its plan. The minimum unsatisfiable core is also provided in the retry prompt as an explanation for the tool call being blocked. To prevent infinite retry loops and to account for potential policy encoding or solver failures, the agent is permitted a bounded number of replanning attempts (up to three in our implementation) before returning control to the user.

\section{Evaluation Setup}

\textbf{Evaluation Setup} We evaluated the solver-aided policy checker on the airline domain of $\tau^2$-bench, which is a benchmark designed to assess tool-augmented agents interacting with users in multi-turn conversations in realistic scenarios. The benchmark consists of domain-specific tasks that require agents to interpret user intents, invoke external tools correctly, and achieve the expected outcome while adhering to complex policy constraints. The airline domain includes $50$ tasks, $13$ tools, and a $1242$-word policy document with detailed rules about cancellations, re-bookings, and compensation eligibility. We specifically select this domain as it contains a more complex domain policy and tasks that can be used to test policy adherence.

We measure task performance with the pass$^{\wedge}$k metric introduced by the benchmark authors. This metric computes the probability that all $k$ i.i.d. task trials are successful, averaged across all tasks. Higher pass$^{\wedge}$k scores indicate increased consistency and reliability across $k$ independent trials.

For all experiments, we run the benchmark $k = 4$ trials to produce pass$^{\wedge}1\ldots4$ scores. The customer and service agents were instantiated with GPT-4.1, and tool call validation used GPT-4o.

\textbf{Research Questions} Specifically, we aim to answer the following research questions:

\begin{itemize}
    \item RQ1: How effective is automated translation of natural-language policies into SMT constraints??
    \item RQ2: How much does solver-aided policy checking reduce the number of policy-violating tool calls made by TaLLM agents?
    \item RQ3: How does solver-aided policy checking impact overall task accuracy and consistency?
\end{itemize}

\section{Results}

\begin{figure*}[t]
  \centering
  \begin{minipage}[t]{0.49\linewidth}
    \centering
    \includegraphics[width=\linewidth]{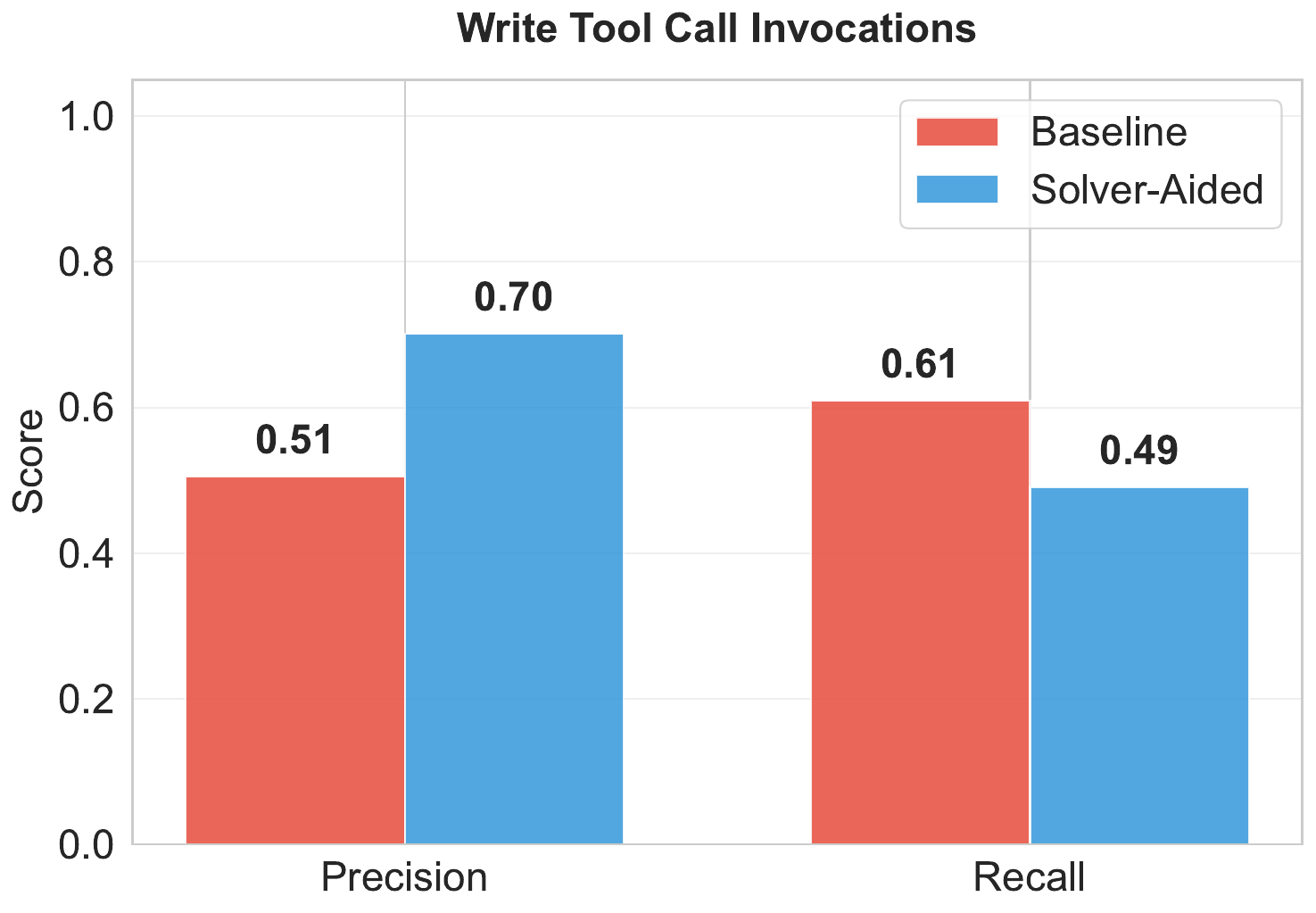}
    \caption{Precision and recall of write tool calls for baseline \tallm and the \tallm with policy checker.}
    \label{fig:policy-violation-proportion}
  \end{minipage}\hfill
  \begin{minipage}[t]{0.49\linewidth}
    \centering
    \includegraphics[width=\linewidth]{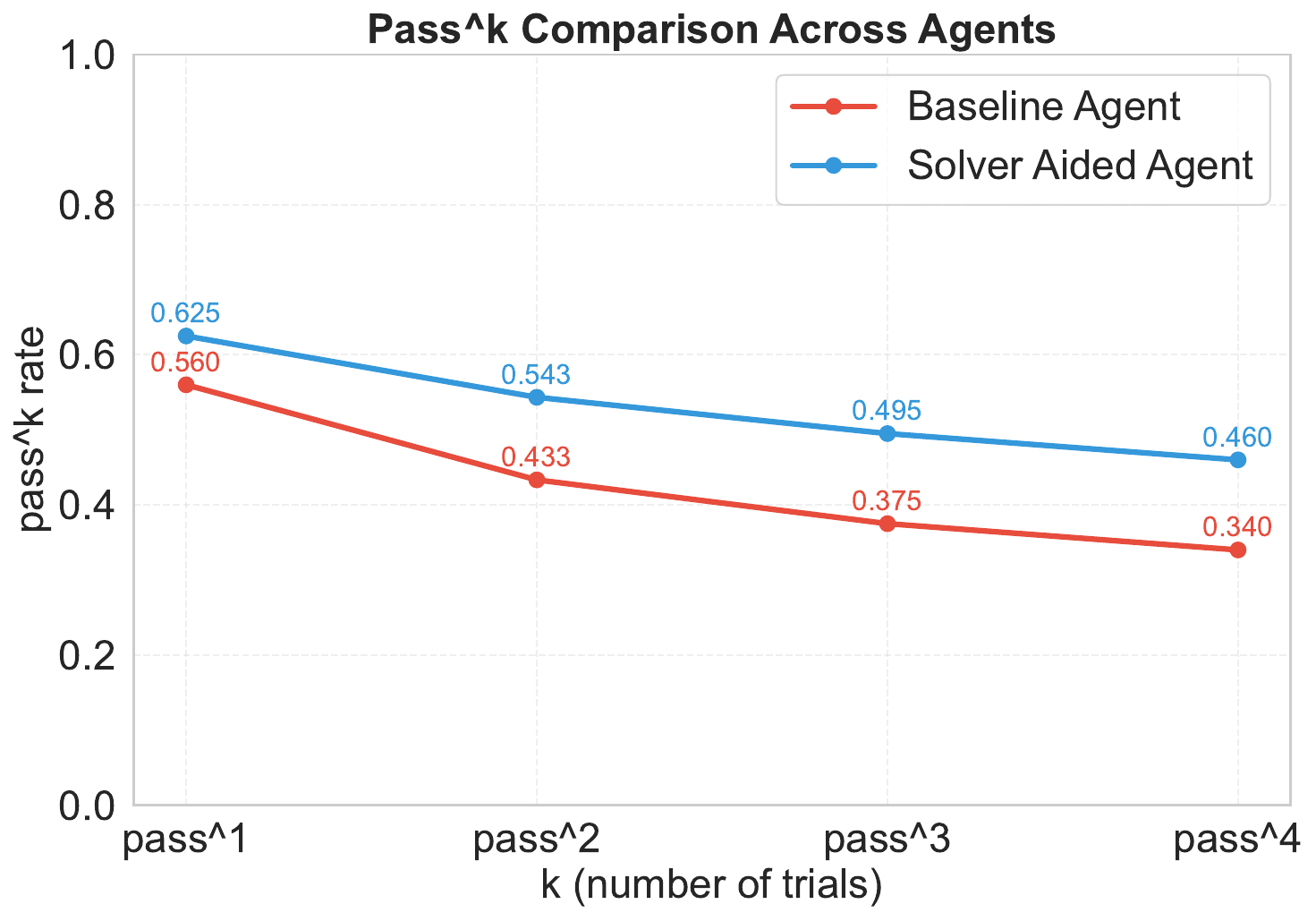}
    \caption{The pass$^{\wedge}$k scores for the baseline \tallm and the \tallm with policy checker.}
    \label{fig:pass-k-lines}
  \end{minipage}
\end{figure*}

\subsection{RQ1: Formal Logic Policy Translation}

We investigated several approaches for encoding natural-language policies into SMT constraints, progressing from fully automated generation to human-in-the-loop methods. Across these designs, we observed recurring challenges in syntactic correctness, semantic completeness, and constraint tightness. Below, we summarize the designs explored and the outcomes observed.

\textbf{Design 1: Proof-of-Thought (Simple LLM) Generation.} Our initial attempt relied on the Proof-of-Thought framework \cite{ganguly2024pot}, which involved prompting an LLM to directly generate SMT-LIB and executing the generated code with Z3. This approach consistently failed on realistic policy documents: the generated SMT-LIB frequently contained syntax errors, undefined symbols, and incomplete rule structures. As a result, this design was not viable for policy verification.

\textbf{Design 2: Iterative LLM Generation and Repair.} Our next approach built on top of the Proof-of-Thought framework by adding in an iterative repair loop. The LLM generated an initial encoding, and errors produced by Z3 when running the generated code were passed back to the LLM for further refinement and repair. This approach produced syntactically correct SMT-LIB (50–60 lines) and performed better when generating Z3-Python code than raw SMT-LIB. However, the major limitation of this approach was that the resulting policy encodings omitted many essential constraints, leading to underconstrained policies and incorrect behavior.

\textbf{Design 3: Automated Reasoning Tool in AWS Bedrock.} We next applied an automated-reasoning-capable LLM via the Automated Reasoning tool in AWS Bedrock. This approach produced more complete policy encodings (around 600 lines) with around 95\% coverage according to an LLM judge. While significantly improved, many rules remained underconstrained. For example, the encoding for the cancellation policy used an implication to allow cancellations when a booking was made within 24 hours, but did not explicitly prohibit cancellations outside this window. As a result, the solver permitted tool invocations that violated the intended policy semantics.

\textbf{Design 4: AWS Bedrock + Custom Prompting + Manual Tuning.} Due to the limitations of using AWS Bedrock out of the box, the final design augmented the Bedrock approach with custom prompting instructions to emphasize constraint correctness tightness. We further performed manual review and targeted tuning on the generated SMT-LIB to refine the rules. Some of these manual improvements included adding definitions for undefined variables, adding timestamp conversions for time-based constraints, reducing complexity in variable declarations, and diagnosing underconstrained implications by running the benchmark and repairing them. This process yielded the final translated SMT-LIB policy encoding that we used for the \tallm.

\textbf{Observed Challenges.}
Across all designs, three core challenges emerged:
\vspace{-2pt}
\begin{itemize}
    \item \textbf{Syntactic Correctness.} Automatically generated policy encodings frequently fail to have correct syntax and typing requirements.
    \item \textbf{Semantic Completeness.} Even syntactically valid encodings often omit policy clauses or fail to capture more complex policy interactions.
    \item \textbf{Constraint Tightness.} Natural-language policies commonly rely on implicit negation and exclusivity assumptions. Failure to make this explicit results in under-constrained models that allow policy-violating tool calls.
\end{itemize}

These findings indicate that automatically translating natural-language policies into logical constraints is a challenging task. Furthermore, verifying semantic completeness requires manual review. As a result, we treat policy encodings as explicit, reviewable artifacts rather than relying solely on fully automated translation. Future work will be needed to develop a fully automatic encoding of policies.


\subsection{RQ2: Reduction in Policy-Violating Tool Calls}

To evaluate the effectiveness of solver-aided policy checking in preventing policy violations, we ran the $\tau^2$ benchmark with and without the policy checker enabled. Figure~\ref{fig:policy-violation-proportion} shows the number of expected write tool calls across the four trials of all the tasks as well as the number of valid and invalid tool calls that were made by the baseline \tallm and the \tallm with policy checker.

While the baseline produced a roughly equal proportion of valid and invalid write tool calls, the \tallm with policy checker reduced the fraction of invalid write tool calls to $29\%$ (measured over all write tool calls). However, it also produced fewer correct write tool calls overall compared to the baseline. This suggests a tradeoff: policy checking improves precision (fewer invalid tool calls) at the cost of a slight reduction in recall (fewer correct tool calls).

  


\subsection{RQ3: Impact on Overall Task Accuracy and Consistency}

We measured task performance using the pass$^k$ metric, which evaluates consistent task success across $k$ independent executions of the same task. Figure~\ref{fig:pass-k-lines} compares the \tallm with policy checker to the baseline \tallm across all tasks in the airline domain for $k = 4$. Overall, both achieve similar success rates, with the policy checker yielding a slight improvement. Moreover, the policy checker improves consistency across runs: its success rate decreases by only $26\%$ as $k$ increases, compared to a $40\%$ decrease for the baseline.

\section{Related Work}

\subsection{Tool Augmented LLMs}

Recent work has extended the capabilities of language models beyond text generation by enabling LLMs to use external, domain-specific tools. These Tool-augmented LLM (TaLLM) agents vary in the way they provision tools to the LLM and execute them \cite{lu2023chameleon, patil2023gorilla, hao2023toolkengpt, dam2024survey}. For example, Chameleon \cite{lu2023chameleon} provides documentation for compositional tools in the LLM system prompts to accomplish question-answering tasks, while Gorilla \cite{patil2023gorilla} is a fine-tuned LLM that selects machine learning tools for a task.

\subsection{Policy Enforcement in LLM Agents}

Recent work has aimed to characterize the types of failures that TaLLM agents exhibit, either by developing new benchmarks to analyze various task capabilities on different domains \cite{li2023apibank, huang2024planning, webarena2024, taskbench2024, metatool2024, yao2024taubench} or by more systematic analysis of existing benchmarks \cite{winston2025taxonomy}. This work has revealed common failures, including incomplete task planning, incorrect tool invocations, and tool execution errors.

Policy adherence is a key failure mode, as highlighted by Rodriguez et al., 2025 \cite{rodriguez2025policybench} and Tau-bench \cite{yao2024taubench}. Existing LLM agents primarily enforce policies by including such domain-specific rules in the system prompt or by filtering or modifying responses in a post-processing step. However, these approaches lack deterministic guarantees of correctness and cannot prevent policy-violating actions from being executed.

\subsection{Formal Verification and SMT Solvers}

On the other hand, formal methods can provide hard guarantees for enforcing policy constraints in LLM agents. A recent framework (Proof of Thought \cite{ganguly2024pot}) compiles natural language reasoning into logical proofs checked by solvers. Yet, no existing work integrates such formal verification directly into the tool-use planning loop of the LLM. Our work bridges this gap by integrating proof-based reasoning into the action planning loop of tool-using agents.

\section{Conclusion}

We presented a solver-aided policy checker framework for enforcing tool-call policy compliance in TaLLM agents. By compiling natural-language operational policies into logical constraints and checking proposed tool invocations against these constraints at runtime, our approach provides policy enforcement independent of agent reasoning variability. Evaluations on the $\tau^2$ benchmark show that solver-aided checking reduces the frequency of policy-violating tool calls and improves consistency across repeated executions.

Our results highlight both the feasibility and the limitations of applying formal methods to TaLLM policy enforcement. While manual or semi-structured policy encoding remains necessary to ensure constraint correctness and tightness, the resulting encodings can be applied for runtime policy enforcement.

This work suggests several directions for future research, including further automation of policy encoding and improved policy checking. Overall, our work suggests the importance of providing formal guarantees on safety-critical properties, such as domain-specific policies, in tool-augmented LLM agents.

\bibliographystyle{ACM-Reference-Format}
\bibliography{references}

@article{lu2023chameleon,
  title={Chameleon: Plug-and-Play Compositional Reasoning with Large Language Models},
  author={Lu, Pan and Peng, Baolin and Cheng, Hao and Galley, Michel and Chang, Kai-Wei and Wu, Ying Nian and Zhu, Song-Chun and Gao, Jianfeng},
  journal={arXiv preprint arXiv:2304.09842},
  year={2023},
  url={https://arxiv.org/abs/2304.09842}
}

@article{patil2023gorilla,
  title={Gorilla: Large Language Model Connected with Massive APIs},
  author={Patil, Shishir G. and Zhang, Tianjun and Wang, Xin and Gonzalez, Joseph E.},
  journal={arXiv preprint arXiv:2305.15334},
  year={2023},
  url={https://arxiv.org/abs/2305.15334}
}

@article{li2023apibank,
  title={API-Bank: A Comprehensive Benchmark for Tool-Augmented LLMs},
  author={Li, Moxin and Zhao, Yuxin and Yu, Boyuan and Song, Fuwen and Li, Hongyang and Yu, Hao and Li, Zhifei and Huang, Fei and Li, Yulong},
  journal={arXiv preprint arXiv:2311.04897},
  year={2023},
  url={https://arxiv.org/abs/2311.04897}
}

@article{hao2023toolkengpt,
  title={ToolkenGPT: Augmenting Frozen Language Models with Massive Tools via Tool Embeddings},
  author={Hao, Sheng and Liu, Tianyang and Wang, Zhiyu and Hu, Zhiting},
  booktitle={Thirty-seventh Conference on Neural Information Processing Systems},
  year={2023}
}

@article{dam2024survey,
  title={A Complete Survey on LLM-based AI Chatbots},
  author={Dam, S. K. and Hong, C. S. and Qiao, Y. and Zhang, C.},
  journal={arXiv preprint arXiv:2406.16937},
  year={2024},
  url={https://arxiv.org/abs/2406.16937}
}

@article{huang2024planning,
  title={Planning, Creation, Usage: Benchmarking LLMs for Comprehensive Tool Utilization in Real-World Complex Scenarios},
  author={Huang, Shijue and Zhong, Wanjun and Lu, Jianqiao and Zhu, Qi and Gao, Jiahui and Liu, Weiwen and Hou, Yutai and Zeng, Xingshan and Wang, Yasheng and Shang, Lifeng and Jiang, Xin and Xu, Ruifeng and Liu, Qun},
  journal={arXiv preprint arXiv:2401.17167},
  year={2024},
  url={https://arxiv.org/abs/2401.17167}
}

@article{taskbench2024,
  title={TaskBench: Benchmarking Large Language Models for Task Automation},
  author={Shen, Yongliang and Song, Kaitao and Tan, Xu and Zhang, Wenqi and Ren, Kan and Yuan, Siyu and Lu, Weiming and Li, Dongsheng and Zhuang, Yueting},
  journal={arXiv preprint arXiv:2311.18760},
  year={2024},
  url={https://arxiv.org/abs/2311.18760}
}

@article{metatool2024,
  title={MetaTool Benchmark for Large Language Models: Deciding Whether to Use Tools and Which to Use},
  author={Huang, Yue and Shi, Jiawen and Li, Yuan and Fan, Chenrui and Wu, Siyuan and Zhang, Qihui and Liu, Yixin and Zhou, Pan and Wan, Yao and Gong, Neil Zhenqiang and Sun, Lichao},
  journal={arXiv preprint arXiv:2310.03128},
  year={2024},
  url={https://arxiv.org/abs/2310.03128}
}

@article{webarena2024,
  title={WebArena: A Realistic Web Environment for Building Autonomous Agents},
  author={Zhou, Shuyan and Xu, Frank F. and Zhu, Hao and Zhou, Xuhui and Lo, Robert and Sridhar, Abishek and Cheng, Xianyi and Ou, Tianyue and Bisk, Yonatan and Fried, Daniel and Alon, Uri and Neubig, Graham},
  journal={arXiv preprint arXiv:2307.13854},
  year={2024},
  url={https://arxiv.org/abs/2307.13854}
}

@inproceedings{winston2025taxonomy,
  title={A Taxonomy of Failures in Tool-Augmented LLMs},
  author={Winston, Cailin and Just, René},
  booktitle={2025 IEEE/ACM International Conference on Automation of Software Test (AST)},
  year={2025},
  month={April 28-29},
  publisher={IEEE},
  url={https://ieeexplore.ieee.org/document/11081716}
}

@article{yao2024taubench,
  title={Tau-Bench: A Benchmark for Tool-Agent-User Interaction in Real-World Domains},
  author={Yao, Shunyu and Shinn, Noah and Razavi, Pedram and Narasimhan, Karthik},
  journal={arXiv preprint arXiv:2406.XXXX}, 
  year={2024}
}

@inproceedings{ganguly2024pot,
  title={Proof of Thought: Neurosymbolic Program Synthesis Allows Robust and Interpretable Reasoning},
  author={Ganguly, Debargha and Iyengar, Srinivasan and Chaudhary, Vipin and Kalyanaraman, Shivkumar},
  booktitle={38th Conference on Neural Information Processing Systems (NeurIPS 2024) System 2 Reasoning At Scale Workshop},
  year={2024}
}

@article{rodriguez2025policybench,
  title={Towards Safer Chatbots: A Framework for Policy Compliance Evaluation of Custom GPTs},
  author={Rodriguez, David and Seymour, William and Del Alamo, Jose M. and Such, Jose},
  journal={arXiv preprint arXiv:2502.01436},
  year={2025},
  url={https://arxiv.org/abs/2502.01436}
}

\end{document}